\documentclass[prl,twocolumn,showpacs,amsfonts,amssymb]{revtex4}

\usepackage{graphicx}
\usepackage{amsmath}
\usepackage{bm}
\usepackage{times}
\newlength{\figwidth}
\renewcommand{\abstractname}{}\abstractname
\begin{document}
\title{Demonstration of perfect coherence preservation for matter-waves perturbed by a $\delta$-kicked rotor}
 
\author{A. Tonyushkin}
\email{alexey@physics.harvard.edu}
\author{S. Wu}
\altaffiliation[Present address: ]{NIST, Gaithersburg, MD, 20899}
\author{M. Prentiss}
\affiliation{Department of Physics, Harvard University, Cambridge, MA 02138}
\begin{abstract}
We demonstrate perfect coherence preservation in an atom interferometer perturbed by kicks from off-resonant standing wave pulses. Under most conditions, the decoherence induced by the pulses reduces the signal; however, the coherence is perfectly preserved when the kicking period is equal to the rational fraction of the inverse atomic recoil frequency, independent of the number or the randomness of the strength of the applied kicks. The width narrowing of coherence revival as a function of increasing kick number and strength provides a new accurate measurement of the recoil frequency.
\end{abstract}
\pacs{3.75.Dg, 32.80.Pj, 05.45.Mt, 39.20.+q}
\maketitle 

Coherence and interference are fundamental properties of quantum mechanical systems.  
Internal state coherence has played a vital role in enhancing our basic understanding of the universe~\cite{EPR}, and underlies widely used devices such as atomic clocks, lasers, and MRI scanners. External state coherences in the electron gas result in the very intriguing phenomenon of Anderson localization~\cite{Anderson} and are increasingly important in experimental physics with the advent of low dimensional electron gas and gaseous Bose-Einstein condensate (BEC) in disordered potentials~\cite{2DEG}. 
The external state coherence of lasers provides the basis for many inertial sensing systems, and it is hoped that the interference of external atomic states will provide even more sensitive and compact sensing systems~\cite{Wu}.
While the sensitivity of the coherences to the environment is an advantage for sensing, interactions that destroy coherences limit many experiments and greatly hinder the development of interference based devices~\cite{Decoherence}, including quantum computers~\cite{Internal}. Fortunately, not all coupling to the environment results in decoherence.  For example, it has been shown that the internal state coherence between different magnetic sublevels can be preserved even in the presence of a magnetic field, as long as the magnetic field shifts for both states are the same~\cite{Cornell}.  Similarly, photon gyroscopes exploit reciprocal paths so that time independent perturbations in the phase of the external states cancel~\cite{Gyro}. 

In this paper, we will show that external state coherences between different momentum states can be perfectly preserved in an atom interferometer that is strongly perturbed by a pulsed spatially dependent potential, even when the potential is applied more than eighty times and successive pulses have random strengths. In particular, we use a pulsed off-resonant optical standing wave (SW) to create a periodic spatially dependent potential that approximates delta kicks in the atomic $\delta$-kicked rotor (ADKR) implementation of a quantum kicked rotor~\cite{Raizen95}.  

Previous work has shown that under some circumstances the ADKR initially produces a decoherence, but that successive kicks can produce a revival and finally saturation of the coherence at a finite value in the limit where the number of kicked rotor pulses approaches infinity~\cite{Badtime}. Our results show that under certain conditions successive applications of the ADKR produce no decoherence at all, even when the ADKR significantly changes the momentum distribution of the atoms. This remarkable experimental result shows that strong perturbation need not produce any decoherence and demonstrates a point of practical importance: external state coherence can be perfectly preserved in pulsed optical lattices; therefore, the decoherence observed in other interferometers interacting with pulsed potentials cannot simply be attributed to the interaction with the pulsed lattice~\cite{Wang05}.

\begin{figure}[htb!]
\begin{center}
\includegraphics[width=\figwidth]{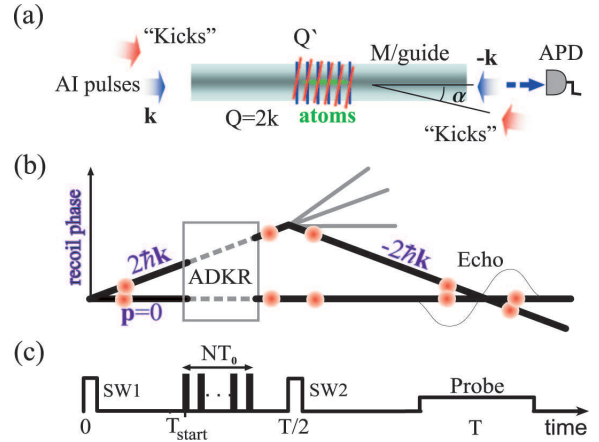} %
\end{center}
\caption{(color online) Experimental setup: (a) configuration along the magnetic guide;
(b) matter-wave recoil diagram; and (c) a pulsing scheme. Here, AI = atom interferometer, M/guide = magnetic guide for atoms, APD = avalanched photo-diode, ADKR = atomic $\delta$-kicked rotor, SW1,2 = 1st and 2nd standing waves of AI.}
\label{fig:experiment}
\end{figure}

In our experiment we subject a cloud of cold $^{87}$Rb atoms in a magnetic guide to periodic kicks from a sinusoidal potential created by SW pulses of off-resonant laser light along the guiding direction. 
The experimental setup and a pulsing scheme are shown in Fig.~\ref{fig:experiment}.
%
%
We realize the ADKR model potential by applying a train of off-resonant SW pulses in between the two interferometer pulses (SW1 and SW2 in Fig.~\ref{fig:experiment}(c)), therefore we treat a $\delta$-kicked rotor as a perturbation to the matter-wave's dynamics in a de~Broglie wave interferometer.  The long trapping time and corresponding long coherence time of the atom interferometer in a magnetic guide~\cite{Wu} allow us to investigate ADKR for the large number of pulses $N$ and over a broad range of kicking strengths, characterized by the pulse area $\theta$. 
For sufficiently short kicks, the motion of atoms during pulse duration $t_p \approx 0.5\, \mu$s, which is a Raman-Nath regime, can be neglected . Also for a pulse duration much less than the kick period $T_0$ one can approximate each kick as a delta function~\cite{Raizen99}. The kicking pulses are 6.8~GHz red detuned from  $F=1$ ground hyperfine state of $^{87}$Rb and the maximum achievable pulse area $\theta \approx 9.5$ as estimated from comparison of experimental signal to simulations.
As shown in Fig.~\ref{fig:experiment}(a), the propagation direction of the $\delta$-kicked rotor SW is tilted by $\sim 40 $~mrad angle with respect to the interferometer SW. 
The grating vectors $Q$ and $Q^{\prime}$ of the respective interferometer and kicked rotor SW fields are not equal.
Our detection method relies on coherent backscattering of the probe pulse from the atomic grating induced by the interferometer pulses; therefore, it is only sensitive to spatial distributions with the periodicity of the interferometer SW $2\pi/Q$, which is not equal to $2\pi/Q^{\prime}$. Thus, we are not sensitive to the spatial interference patterns due to the ADKR, and the ADKR can simply be considered as a perturbation to the phase evolution of the interferometer signal. 
A detailed description of the de~Broglie wave interferometer in a magnetic guide is given elsewhere~\cite{Wu}. 
The interferometer SW consists of the two counter-propagating traveling waves with k-vectors ${\bf k}$ precisely aligned along the guide and the SW field is $\sim 20\Gamma$ blue detuned from $F=1 \!\rightarrow F^\prime =2$ transition. The schematic of matter-waves diffraction in the atom interferometer is conveniently described by the recoil diagram~\cite{Cahn} in Fig.~\ref{fig:experiment}(b).
 
\begin{figure}[htb!]
\begin{center}
\includegraphics[width=\figwidth]{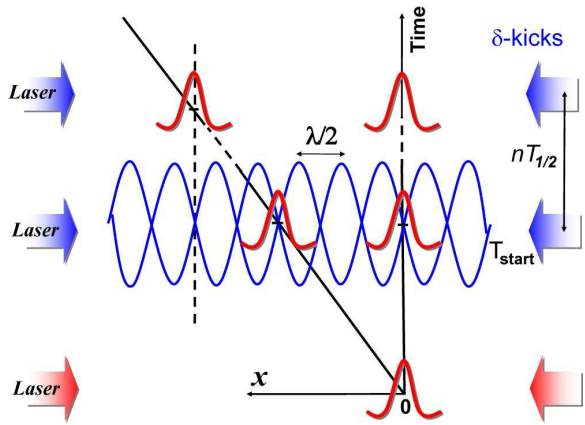} %
\end{center}
\caption{(color online) A schematic diagram of wave packet dynamics with coherence preservation. At time $t=0$ the wave packet (shown in red) is diffracted by the interferometer SW pulse into two wave packets which at time $T_{start}$ interact with kicking SW (shown in blue).}
\label{fig:robust}
\end{figure}

The interaction of a kicked rotor SW is schematically shown in Fig.~\ref{fig:robust}. 
%
%
The atomic wave packet is split by the interferometer SW pulse at time $t=0$ into two wave packets with momenta that differ by $\hbar Q$. The interaction with the kicked rotor potential begins at a time $t=T_{start}$ when the first kicking SW with a grating vector $Q^{\prime}$ is applied. 
If the spatial displacement between two diffracted wave packets at the time of interaction is commensurate with the spatial period of the kicking SW pulse then both wave packets experience the same potential; therefore, the differential phase between those wave packets is equal to zero and the interferometer signal is completely unaffected by this interaction. The same argument applies for interactions at times \{$T_{start}+ n T_{1/2}$\}, where $n$ is an integer, $T_{1/2} = \pi /\omega _{Q^{\prime}}$ is the {\em half-Talbot time} ($T_{1/2} \approx 33.2 \, \mu$s for $^{87}$Rb), and $\omega _{Q^{\prime}} = \hbar Q^{\prime 2} /2m$ is a two-photon atomic recoil frequency for the kicked rotor potential. 
Contrary to previously observed quantum kicked rotor resonances~\cite{QR} that require regular spacing and strength of the applied SW pulses, in our case the invariance under the ADKR perturbation does not rely on the temporal pattern of the kicks and is still valid for the infinitely large single kick or random strength kicks separated by irregular integer multiples of $T_{1/2}$. To distinguish between general quantum resonance conditions we call the latter a {\em robust resonance}. 
The theoretical description of ADKR impact on the atomic coherences is presented elsewhere~\cite{Badtime}. Here we present the result for the robust resonance case. 
The normalized coherence signal, given by the Fourier component of the atomic density grating $f=\rho_{-Q} ^{V}(T) / \rho_{-Q}(T)$, can be calculated according to the expression~\cite{Badtime}
%
\begin{equation}
	f=\sum_{\{n_i\}} \prod_{i=1} ^N J_{n_i}\left( 2\theta \sin{\frac{Q^{\prime} X^{n_i} _i}{2}}\right) \,,
\label{Coherence}
\end{equation}
%
where $\theta$ is the pulse area of a kicked rotor potential, $\{ n_i \}$ defines the set of all the contributing diffraction paths due to interaction with the kicked rotor, $X^{n_i} _i$ is the spatial displacement between the atomic wave packets for a diffraction vortex set at fixed time. For the kicked SW with the period $n T_{1/2}$ the spatial displacement is given by 
$X(T_{start})= \delta x(T_{start}) + 2n \pi /Q^{\prime}$, where $\delta x(T_{start})$ is the initial spatial displacement. The robust resonance condition introduced above (also see Fig.~\ref{fig:robust}) corresponds to the initial spatial displacement which is commensurate to the period of the kicking SW $2\pi/Q^{\prime}$. Therefore, we can write the displacement as $ X(T_{start}^{res})= 2(n+m)\pi/Q^{\prime}$ for $n , m$ integers. 
The differential strength defined as 
$\phi = 2\theta \sin{ Q^{\prime} X(T_{start})/2}$~\cite{Badtime} is equal to zero for such a condition; therefore, in Eq.(\ref{Coherence}) only $0^{th}$-order Bessel functions contribute to the sum of all the diffraction paths which proves the invariance of the coherence signal under any number of kicks independent on the perturbation strength $\phi$ at any given kick.

\begin{figure}[htb!]
\begin{center}
\includegraphics[width=\figwidth]{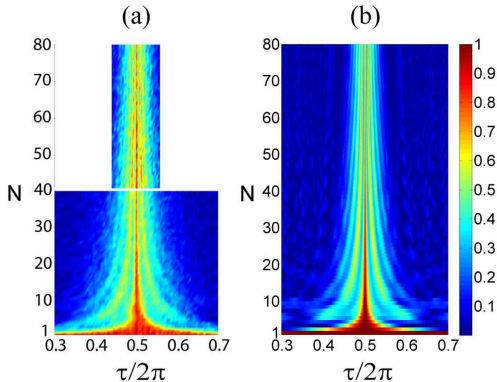} %
\end{center}
\caption{(color online) Matter-wave coherence dynamics: a) experimental data and b) corresponding simulations ($\theta =0.7$). Here, the horizontal axis is the normalized kicking period, the vertical axis is the number of kicks. The data show the perfect normalized coherence preservation (red color on the contour plot) in the narrow vicinity of the kicking period $\tau=\pi$ ($T_0 \approx 33.2 \, \mu$s). (Note that there is a different resolution used for experimental and simulation plots. Interpolation is used between the experimental data points.)}
\label{fig:1stecho}
\end{figure}

\begin{figure}[htb!]
\begin{center}
\includegraphics[width=\figwidth]{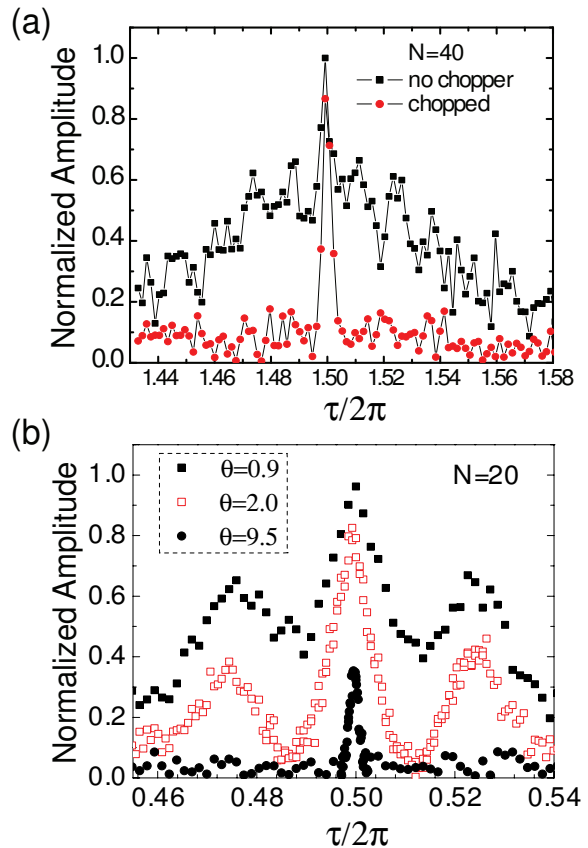} %
\end{center}
\caption{(color online) Robustness of coherence preservation: (a) coherence amplitude at $N=40$ vs kicking period corresponding to the regular (black squares) and random (red circles) kicking patterns (the curves are used to guide the eye); (b) resonance peak at finer time scale for $N=20$ and $\theta=0.9, 2.0, 9.5$.}
\label{fig:noise}
\end{figure}

Figure~\ref{fig:1stecho} shows the effect of the ADKR on the coherence between the momentum states that differ by $2 \hbar k$ for up to $N=80$ kicks applied and the pulse area of the kicking SW $\theta =0.7$, as determined by the simulations fit. The experimental data are given in Fig.~\ref{fig:1stecho}(a) where for each number of kicks {\em N} we measured the backscattered echo signal at time $T$ as a function of the normalized kicking period $\tau =\omega _{Q^{\prime}}T_0$. In practice $\omega _Q - \omega _{Q^{\prime}} \ll \omega _Q$, so for convenience we ignore the difference and use $T_{1/2} = \pi / < \omega _{Q^{\prime}}> = 33.2 \, \mu$s. The experimental scan as a function of the kicking period consists of two parts: coarse for $N=1-40$ and fine for $N=41-80$, where for greater detail we scanned the kick period with higher resolution in the narrow vicinity of the resonance $\tau /2\pi \in \left[0.44,0.55\right]$. The corresponding simulations are shown in Fig.~\ref{fig:1stecho}(b). 
%
%
The value of the coherence is represented by false color scaled according to the color-bar where red corresponds to unity. It is seen from the data that the coherence gets depleted everywhere except a narrow vicinity of the resonance.
The delay time between the interferometer SW pulse and the first kicking pulse $T_{start}$ is chosen so that the separation
between the two atomic wave-packets is an integer multiple of $\pi/Q^{\prime}$. We precisely determine the required $T_{start}$ which corresponds to a maximum of an echo amplitude from the recoil oscillations experiment where the echo amplitude is plotted versus the delay of the kicking pulse.  
In agreement with the qualitative picture given in Fig.~\ref{fig:robust}, both simulations and experiment show that the lowest order coherence that is normalized to the unperturbed interferometer
signal is perfectly preserved for the kicking period equal to the first ``resonance'' condition $\tau=\pi$.
The observed finer fringes near the resonance coherence peak analogous to the fringes in a double slit diffraction pattern. 
If the kicking strength is sufficiently small, these fringes produce high background around the sharp robust resonance peak; however, at longer kicking period $\tau$ (experimentally accessible $37\pi$) decoherence mechanisms associated with gravity or stray magnetic fields remove the background. Even for $\tau \sim \pi$, we will show that the background can be removed by either applying high area optical pulses or by randomizing the strengths of successive lower power pulses~\cite{noise}. The removal of the background by applying random amplitude kicks generated by adding an optical chopper to randomly block the kicked rotor light is shown in Fig.~\ref{fig:noise}(a), where the kicking period was scanned near the third resonance $\tau=3\pi$ and the chopping period was of the same order but incommensurate with it. Though the background is removed, the coherence of the robust central resonance is preserved. The removal of the background by higher strength pulses is illustrated in Fig.~\ref{fig:noise}(b), where the number of identical kicks is fixed at $N=20$ but the pulse areas are changed. At the highest pulse areas, the coherence loss due to spontaneous emission becomes important, resulting in a decrease in the size of the robust resonance. 
%

\begin{figure}[htb!]
\begin{center}
\includegraphics[width=\figwidth]{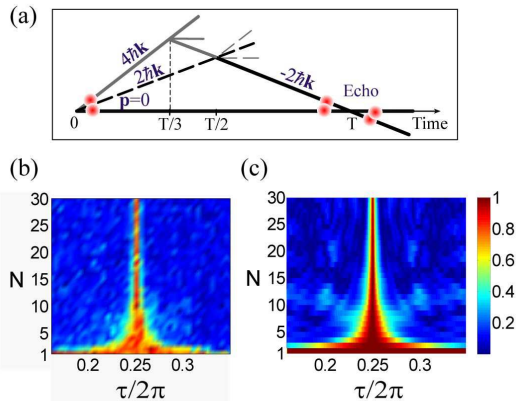} %
\end{center}
\caption{(color online) Effect of ADKR on higher order coherence: a) the recoil diagram for the 2nd echo scheme as compared to the 1st echo scheme (dashed line); b) experimental data; c) simulations ($\theta =1$).  On the contour plots the horizontal axis is the normalized kicking period, the vertical axis is the number of kicks.}
\label{fig:2ndecho}
\end{figure}

In general, only the coherences with momentum difference $2\hbar k$ are considered in the so-called ``first echo'' scheme, as shown in Fig.~\ref{fig:experiment}(b) where interferometer pulses are applied at times $t=0$ and $T/2$. 
Second order coherences were observed indirectly in Ref.~\cite{Strekalov} and directly in Ref.~\cite{Mask}.
In our experiment we apply ADKR when the different momentum states are separated in phase space by different amounts, so the detected signal corresponding to the lowest order coherences' harmonic encodes the information of the effect of ADKR on higher order coherences. 
To test that conjecture we also utilized a so-called ``second echo'' scheme~\cite{Cahn} as shown in Fig.~\ref{fig:2ndecho}(a) where the two interferometer SW pulses are applied at times $t=0$ and $T/3$ and the lowest coherences correspond to a $4\hbar k$ momentum difference.  
Here we show the results of such an impact on the 2nd order coherences separated by $4\hbar k$ recoil momenta according to the scheme shown in Fig.~\ref{fig:2ndecho}(a). Figure~\ref{fig:2ndecho}(b),(c) show experimental and simulation data respectively of the 2nd order coherence evolution for up to 30 kicks near the kicking period $\tau/2\pi=0.25$ ($T_0 \approx 16.6 \, \mu$s). 
%
%
We note that for the lowest order coherences ($\hbar Q$) there is no quantum resonance at $\tau=\pi/2$. 
The higher order coherences' dynamics under ADKR perturbation can be probed similarly in the n$^{th}$-echo configuration with the second pulse applied at $t=T/(n+1)$ for integer {\em n}. The corresponding resonances of coherence preservations are expected to occur at integer multiples of $\tau=\pi/n$~\cite{HighOrder}. 
 
The described robust resonance phenomena could be used for precision measurements of atomic recoil frequency $\omega_{Q^{\prime}}$, and hence $\hbar /m_{Rb}$. The precision of such a measurement is approximated by $\Delta\omega_{Q^{\prime}}/\omega_{Q^{\prime}} = \Delta T_{1/2} /(l T_{1/2})$, for an integer $l$. Theoretical analysis of the resonance width shows that it should scale as $\Delta T_{1/2} \sim \delta T_0 /(l \theta N^p)$, where $\delta T_0$ is the width of the resonance after just one kick and $p$ is a decay factor. We experimentally obtained $p=1$ for the lowest order coherences and fixed $\theta$. If we allow $\theta$ to be random during the sequence, a different scaling law is expected and we are currently investigating this regime.
With all current experimental constraints the expected precision is of the order of 10 ppm, which can be further significantly improved by increasing the kicking strength $\theta$ and the number of kicks that can be delivered to the system between the interferometer pulses. The ultimate precision is limited by the total interrogation time of our interferometer. The expected improvement of the precision in our scheme over the previously implemented atom optics-based methods~\cite{Cahn,Recoil} is a factor of $l \theta N$.

In conclusion, we experimentally demonstrated perfect coherence preservation of matter-waves perturbed by a train of optical lattice pulses, when the time between successive pulses is a half integer multiple of the Talbot time. This coherence preservation is independent of the number of kicking pulses applied during the interferometer cycle, and occurs even if the strength of successive pulses is random. The coherence is preserved because there is no differential phase shift associated with the perturbation when the interfering wave-packets traverse a distance of an integer multiple of the spatial period of the kicked rotor potential. For a given interrogation time, increasing the number of kicks narrows the time window during which the coherence is preserved; therefore, kicked rotor increases the accuracy of recoil frequency measurements made using the interferometer.  
    
AT thanks G.~Zaslavsky for stimulating discussions.
We acknowledge the financial support from DARPA under DOD, ONR and the U.S. Department of the
Army, Agreement Number W911NF-07-1-0496, and by the Draper Laboratory. 


\end{document}